\documentclass[%
 reprint,
 amsmath,amssymb,
 aps,
]{revtex4-2}

\usepackage{graphicx}% Include figure files
\usepackage{dcolumn}% Align table columns on decimal point
\usepackage{bm}% bold math
\usepackage[utf8]{inputenc}
\usepackage{amsmath}
\usepackage{amssymb}
\usepackage{bm}
\usepackage{pxfonts}
\usepackage{graphicx}
\usepackage[usenames,dvipsnames]{color}
\usepackage{ulem}
\usepackage{lineno}
\DeclareMathOperator{\Tr}{Tr}

\newcommand{\vect}[1]{\mathbf{#1}}
\newcommand{\vecIII}[3]{\left(\begin{array}{c}#1\\#2\\#3\end{array}\right)}

\newcommand{\smatrixII}[3]{\left(\begin{array}{cc}#1&#2\\#2&#3\end{array}\right)}

\newcommand{\dif}[0]{\mathrm{d}}
\newcommand{\dS}{\dif S}

\newcommand{\inv}[1]{#1 ^ {-1}}
\newcommand{\invs}[1]{#1 ^ {-2}}

\newcommand{\brk}[1]{\left( #1 \right)}
\newcommand{\Brk}[1]{\left[ #1 \right]}
\newcommand{\BRK}[1]{\left\{ #1 \right\}}

\newcommand{\abs}[1]{\left|#1\right|}

\newcommand{\ab}{\bar{\vect{a}}}
\newcommand{\bb}{\bar{\vect{b}}}
\newcommand{\aac}{\vect{a}}
\newcommand{\bac}{\vect{b}}

\newcommand{\Chr}[3]{\Gamma^{#1}_{#2#3}}

\newcommand{\nrml}{\hat{\vect{n}}}

\newcommand{\figref }[1]{Fig.~\ref{#1}}
\newcommand{\Eqref}[1]{Eq.~\ref{#1}}

\newcommand{\chng}[1]{#1}

\begin{document}

\preprint{}

\title{Euclidean frustrated ribbons}% Force line breaks with \\
\author{Emmanuel Si\'{e}fert} \thanks{These authors contributed equally to this work}
\author{Ido Levin} \thanks{These authors contributed equally to this work}
\author{Eran Sharon}
\affiliation{Racah Institute of Physics, The Hebrew University, Jerusalem 91904, Israel}%
\date{\today}
\begin{abstract}

Geometrical frustration in thin sheets is ubiquitous across scales in biology and becomes increasingly relevant in technology. 
Previous research identified the origin of the frustration as the violation of Gauss's \emph{Theorema Egregium}. Such ``Gauss frustration" exhibits rich phenomenology; it may lead to mechanical instabilities, anomalous mechanics and shape-morphing abilities that can be harnessed in engineering systems.
Here we report a new type of geometrical frustration, one that is as general as Gauss frustration.
We show that its origin is the violation of Mainardi-Codazzi-Peterson compatibility equations and that it appears in Euclidean sheets.
Combining experiments, simulations and theory, we study the specific case of a Euclidean ribbon with radial and geodesic curvatures.
Experiments, conducted using different materials and techniques, reveal shape transitions, symmetry breaking and spontaneous stress focusing.
These observations are quantitatively rationalized using analytic solutions and geometrical arguments.
We expect this frustration to play a significant role in natural and engineering systems, specifically in slender 3D printed sheets.
\end{abstract}
\maketitle

%\Ido{What about: Many natural soft-structures, across different scales, have developed the ability to change their shape.}
%Soft shape-shifting structures are abundant in natural systems across scales. 
%Examples range from tendrils~\cite{gerbode12}, pine cones~\cite{dawson97}, seedpods opening~\cite{Armon11}, snapping of carnivorous plants~\cite{forterre05} to amphibian eggs~\cite{Yoneda1982} and invertebrates~\cite{Armon2018}.
\section{Introduction}
Inspired by natural systems~\cite{dawson97,forterre05,gerbode12,Armon11,Yoneda1982}, shape-morphing materials have been extensively developed in the last decades. 
They are believed to be at the core of applications in soft robotics~\cite{rus15}, minimally invasive surgery~\cite{cianchetti2014soft} and architecture~\cite{menges18}, because specific functions based on the structure deformation, such as gripping, lifting, sensing or mobility may be incorporated within the material itself as a response to stimuli~\cite{white20}. These techniques include chemical or temperature swelling \cite{Klein07,Kim12,boley19}, pneumatic actuation~\cite{rus15,Siefert_baromorphs2019,Siefert20}, liquid crystal elastomers~\cite{white15,warner19} or dielectric elastomers~\cite{clarke19}. They are all based on basic geometrical principles that link the local distortion fields and the global shape.
Arguably the best-known and most-used shape-morphing strategy is to impose a gradient of strain through the thickness of a flat slender structure to produce spontaneous curvature and subsequent unidirectional bending, such as in the case of bilayer structures ~\cite{stoney09,timoshenko25}.
When this spontaneous curvature is unidirectional and uniform, no geometric frustration emerges since the structure may isometrically bend into a tubular shape. This mechanism is at play in most soft grippers~\cite{shian2015dielectric,devoe1997modeling,shepherd2011multigait}.
Recently, complex shape transformations could be achieved by generating geometrically frustrated sheets~\cite{Klein07}. It was shown that when the spontaneous curvature of a sheet and its in-plane reference geometry (its reference metric) violate  Gauss's \emph {Theorema Egregium}, the sheet does not have any stress-free configuration and its equilibrium shape is determined by a competition between bending and stretching~\cite{Dervaux08,efrati2009elastic}.

%shape transformations could be achieved by controlling the in-plane deformation in slender plates~\cite{Klein07}. Since Gauss' \emph{Theorema Egregium} ties in-plane distances to Gaussian curvature, in order to avoid lateral stretching the plate must buckle into three-dimensional configuration, resulting in bending.Therefore, the structure is geometrically-frustrated (it has no stress-free configurations) and the shape is determined 

\begin{figure}[!ht]
    \centering
    \includegraphics[width=\linewidth]{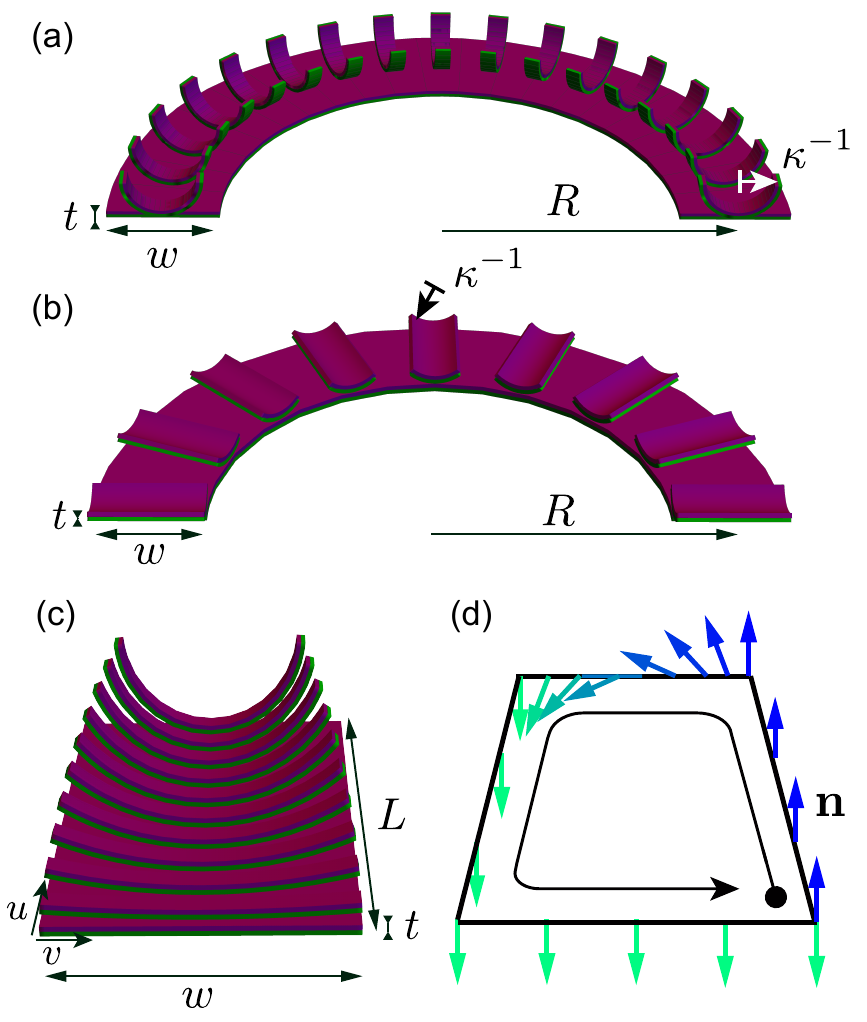}
    \caption{{\bf Configurations of MCP-frustrated ribbons}.
    (a) Curved ribbon with constant radial reference curvature $\kappa$. 
    (b) Curved ribbon with constant azimuthal reference curvature $\kappa$.
    (c) Straight ribbon with linearly varying transverse reference curvature along the longitudinal direction (ranging from $0$ to $\pi/w$).
    (d) Illustration of MCP frustration via the non-conservation of the normal for example (c).
    The normal evolution along the ribbon's boundary \emph{in coordinate space} results in no rotation along the long direction, since there is no reference curvature in this direction. 
    At one end of the ribbon, it rotates by $\pi$ along the width, whereas it does not rotate at the other end.
    Therefore, after integrating the normal evolution over this closed path, we obtain a non-zero rotation of the normal, contradicting the MCP equations, thus leading to an incompatible reference geometry.
    }
    \label{fig1}
\end{figure}

Here, we show that a small variation of the classical bilayers - when the  unidirectional reference curvature slightly varies spatially in amplitude and/or orientation, as shown in Fig.~\ref{fig1} - leads to a previously unstudied class of geometrically frustrated sheets. 
The Gauss equation is indeed satisfied (i.e. the metric is flat and the curvature is everywhere unidirectional) but not the other compatibility condition: the Mainardi-Codazzi-Peterson (MCP) equations.

In this article, we fabricate ribbons whose reference geometry violates the MCP constraints, leading to frustration and pre-stress in the structure.
We show that such simple structures exhibit a rich morphology and present novel scaling laws for the shape transitions.
Such frustrations are expected to appear in many modern manufacturing processes: extrusion \cite{gladman16}, volatilzation \cite{zhang2020rapid} or direct laser writing \cite{bauhofer17} induce a directional shrinkage upon curing: a gradient of residual stresses appears in thin multi-layers structures, causing the spontaneous directional bending of sufficiently thin structures. We also anticipate that this frustration is present and plays a role in biological systems.

\subsection{Incompatible elasticity}
Frustration in slender structures stems from some geometric incompatibilities. On one hand, strain gradients and asymmetry through the thickness induce a \emph{reference curvature tensor} $\bb$. 
On the other hand, mean in-plane deformations - that might be inhomogeneous or anisotropic - result in an intrinsic \emph{reference metric field} $\ab$.
Generally, the elastic energy of isotropic neo-Hookean thin plates and shells may be decomposed into  stretching, $E_S$, and bending, $E_B$, contents and reads~\cite{Ciarlet05}:
\begin{equation}
    \mathcal{E}=\frac{Yt}{8(1-\nu^2)}\int_S \brk{E_S+\frac{t^2}{3}E_B} \,\dS
    \label{eq:energy}
\end{equation}
where $Y$ is the Young modulus, $\nu$ the Poisson ratio, $t$ the thickness and $S$ the surface area.

In the case of geometrically incompatible structures, the bending and stretching contents may be seen as deviations of the \emph{actual geometry} adopted by the structure ($\aac,~\bac$)  from the \emph{reference geometry} ($\ab,~\bb$).
Considering an isotropic Hookean constitutive law, it was shown that the stretching and bending contents may be expressed as~\cite{efrati2009elastic,Dervaux08}:
\begin{align}
 E_S&= \nu \Tr^2\Brk{\ab^{-1}(\aac-\ab)}+(1-\nu) \Tr\Brk{\ab^{-1}(\aac-\ab)}^{2}
 \\
 E_B&=\nu \Tr^2\Brk{\ab^{-1}(\bac-\bb)}+(1-\nu) \Tr\Brk{\ab^{-1}(\bac-\bb)}^{2}
\end{align}

In general, the two reference fields ($\ab,~\bb$) are geometrically incompatible: no surfaces can simultaneously satisfy the reference metric and curvature fields \cite{efrati2009elastic}.
The elastic energy may thus not vanish, but should be minimized, inducing a competition between the bending and stretching contents.
Depending on the relative relevant dimensions of the structure, minimizing the elastic energy is obtained through very different means: when the structure is very thin, the stretching stiffness, which scales linearly with the thickness, is orders of magnitude larger than the bending stiffness, which scales with the cube of the thickness.
The structure thus selects a shape that is an embedding of the reference metric ($\aac=\ab$), canceling the highly unfavourable stretching term at bending cost~\cite{lewicka2011,kupferman2014}.
In this article, this limit is referred to as the \emph{stretching-dominated} regime.
On the other hand, when the structure is sufficiently thick, the frustrated sheet morphs into a shape obeying the reference curvature ($\bac=\bb$) at stretching cost.
We call this limit the \emph{bending-dominated} regime. 
Note that the transition between both regimes does not only involve the thickness but also the other dimensions of the structure, the reference curvatures and the reference Gaussian curvature, transition that we wish to rationalize. 
As we are interested in the minimization of the energy (\ref{eq:energy}) and not in its actual value, we consider and estimate in the following, for the sake of simplicity, the dimensionless energy density denoted as $\mathcal{U}=8(1-\nu^2)/(Y t S)\mathcal{E}$.

\subsection{The Gauss-Mainardi-Codazzi-Peterson equations}
The compatibility equations between the metric and curvature tensors (which completely define a surface up to a rigid-body transformation) are the Gauss-Mainardi-Peterson-Codazzi equations \cite{Ciarlet05}.
 Gauss' equation, known as \emph{Theorema Egregium}, states that the Gaussian curvature, i.e., the product of both principal curvatures, is an intrinsic property of the surface. This means that the Gaussian curvature only depends on distances that are measured on the surface (that is, on the metric $\aac$ alone).
 This can be formulated mathematically as
 \begin{equation}
 K[\aac] = \frac{\det \bac}{\det \aac}    
 \end{equation} 
 
The two MCP equations state that the spatial covariant derivative (the curved space derivative) of the curvature tensor $\bac$ is fully symmetric.
This is so, since $ \aac^{-1}\bac = -\nabla \nrml  $ (where $ \nabla $ is the covariant derivative with respect to $\aac$ and $\nrml$ is the normal to the surface).
Hence, $ \nabla_\alpha\bac_{\beta\gamma} =  \nabla_\beta \bac_{\alpha\gamma} $, which can be formulated as~\cite{struik1961}:
\begin{equation}
    \begin{array}{c}
    \partial_2 L - \partial_1 M = L \Chr{1}{1}{2} + M \brk{\Chr{2}{1}{2}-\Chr{1}{1}{1}} - N\Chr{2}{1}{1}
    \\
    \partial_2 M - \partial_1 N = L \Chr{1}{2}{2} + M \brk{\Chr{2}{2}{2}-\Chr{1}{1}{2}} - N\Chr{2}{1}{2}
    \end{array}
    \label{eq:MCP}
\end{equation}
where $\bac = \smatrixII{L}{M}{N}$ and $\Chr{i}{j}{k}$ are the Christoffel symbols associated with $\aac$.
These compatibility conditions appear as \emph{constraints} when one shapes an actual surface, which geometry is forced to be compatible.
The Gauss equation prohibits for example one from smoothly wrapping a round candy with a flat foil.
Such constraints are also known in origami~\cite{nassar17}, where the MCP equations prohibit the angle of a straight fold from evolving (as a localized counterpart of Fig.~\ref{fig1}c).
Furthermore, curved-fold origami was studied in a series of theoretical works~\cite{dias12,dias2012shape,dias2014non}, and also dealt with these geometrical constraints. 
They appear as a spatially localized version of the example shown in Fig.~\ref{fig1}a.

When considering incompatible elastic sheets, one programs a \emph{reference geometry} and is not limited by these constraints and the sheet's elastic response accommodate for the incompatibility.
This results with shape transitions which are the hallmark of frustrated thin sheets.

\subsection{Mainardi-Codazzi-Peterson frustration}
So far, man-made structures were engineered to be frustrated according to Gauss equation. 
Pioneer shape-morphing materials were programmed with a non-Euclidean reference metric ($K(\ab)\neq 0$) but vanishing curvature ($\bb=0$) \cite{Klein07, Kim12,Siefert_baromorphs2019,white15,clarke19}.
Structures with Euclidean metric ($K(\ab)= 0$) but constant positive ($\det\bb>0$) \cite{Pezzulla16} or negative ($\det\bb<0$) \cite{Armon11,zhang2019shape} reference curvature were also at the core of recent studies.
These systems exhibit rich shape transformations and a sharp transition between  bending-dominated and stretching-dominated regimes.

Frustrated ribbons (long and narrow thin-sheets) are an important subclass of frustrated sheets, that are abundant both in biological~\cite{Armon11} and molecular~\cite{zhang2019shape} systems.
Therefore, a reduced model of \Eqref{eq:energy} was derived~\cite{grossman16, efrati2016non} in which the two reference fields are expanded about the mid-line of the ribbon.
Due to the reduced dimensionality, this has allowed to write analytical models for Gauss-frustrated ribbons and treat thermal fluctuations that dominate frustrated ribbons at the nanometric scale.
However, even in this scope, only frustration resulting from the Gauss equation was studied.

In the case of plates, for which the reference curvature $\bb$ is zero, the MCP equations (\Eqref{eq:MCP}) are trivial. 
Incompatibility may indeed only appear in shells with spatial variation of the curvature.
In Fig. 1, three simple examples of MCP-frustrated ribbons are shown.
The first two correspond to a curved ribbon of geodesic curvature $\kappa_g\equiv 1/R$, width $w$ and thickness $t$ with a unidirectional reference curvature $\kappa$ imposed perpendicularly to (Fig. 1a) or along (Fig. 1b) the ribbon direction.
The third example corresponds to a straight ribbon ($\kappa_g=0$) with a unidirectional reference curvature $\kappa$ that is oriented along the width and varies in the longitudinal direction.
Each ribbon has a Euclidean reference metric and a locally unidirectional reference curvature, such that Gauss equation $K[\ab]=\det\bb/\det\ab=0$ is satisfied. 
However, in all these examples, the MCP-equations are not satisfied.
%These equations ensure that the normal to the surface behaves ``nicely" and recovers its original orientation on any closed loop.
In order to get a geometrical intuition of the MCP violation, one may consider the evolution of the normal on a closed loop in the \emph{coordinate space}: its rotation corresponds to the value of $\bb$ and its parallel transport is given by $\ab$. On an actual surface (thus satisfying the GMPC equations), the normal goes back to its original orientation.
On the other hand, in an MPC-incompatible geometry, the normal does not recover its orientation on a closed loop.
Here, one should not imagine an actual surface (that will trivially obey the MCP equation), but perform this process in coordinate space using $\ab$ and $\bb$.
%One should not imagine an actual surface in 3D space, since we know that no surface having such metric and curvature tensors are immersible in 3D space.
Taking for example the case shown in Fig. \ref{fig1}c, its metric is identically flat (leading to a trivial parallel transport) and its reference curvature reads $\bb=\smatrixII{0}{0}{\pi u/(wL)}$.
Following the evolution of the normal along the boundary (Fig. 1d), we observe that the normal does not return to its initial orientation, since it only rotates by $\pi$ at one end of the ribbon, thus violating the MCP-equations (Fig. 1d).
The same considerations may be applied to the two other examples and also lead to the non-preservation of the orientation of the normal in the coordinate space, indicating MCP frustration.

In this article, we wish to investigate this geometrical frustration, understand the shape selection in the bending and stretching dominated regimes, and characterize the transition between them.
Our study appears as a first step to characterize and understand the shapes obtained and potentially harness directional curvature to engineer target 3D structures from flat printing~\cite{bauhofer17,van2017programming}.

\begin{figure}[!ht]
    \centering
    \includegraphics[width=\linewidth]{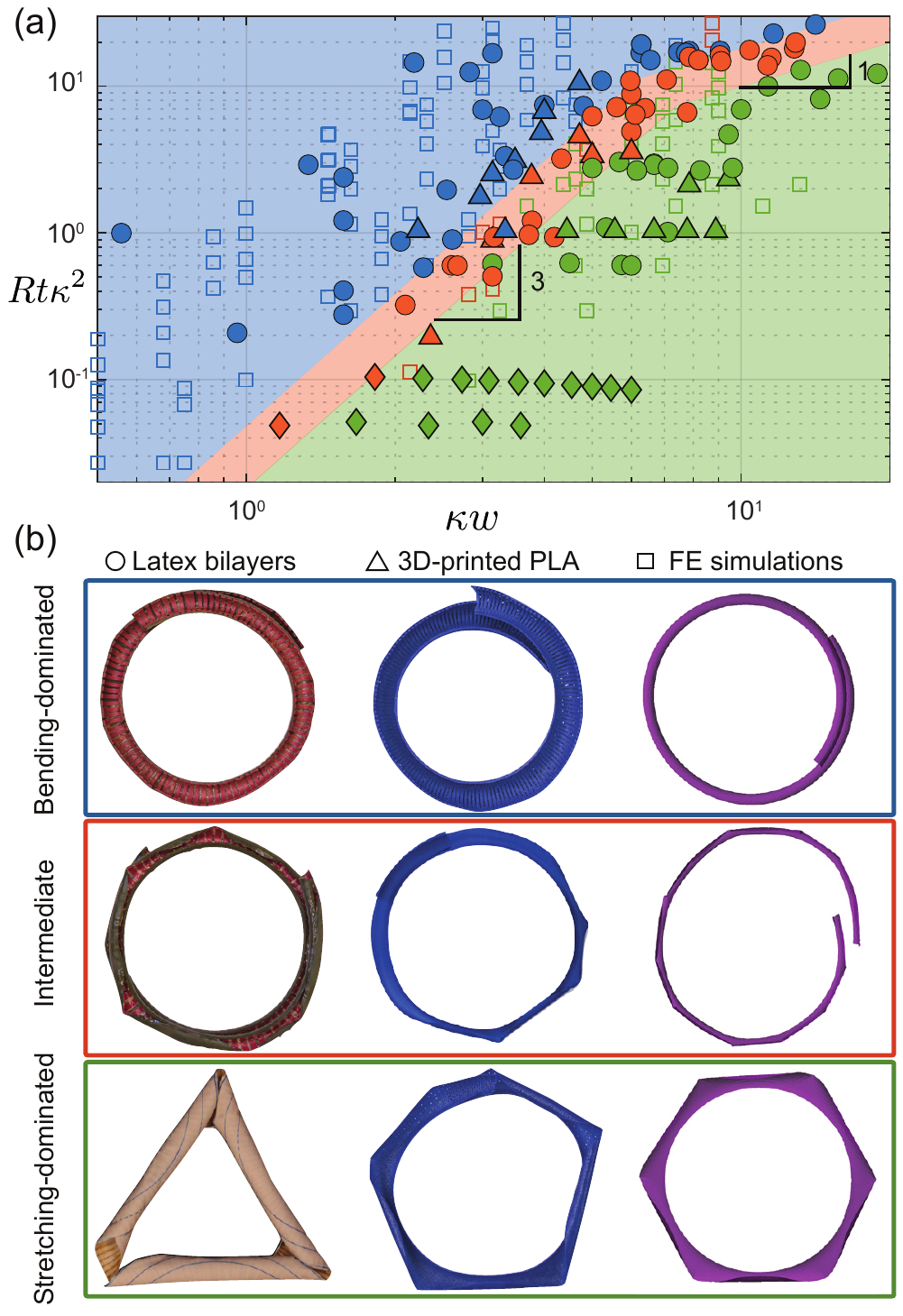}
    \caption{{\bf Morphologies of radially-curved ribbons}. 
    (a) Phase diagram of the three different types of observed shapes: toroidal (blue), intermediate (orange) and faceted (green). 
    Full markers correspond to different experimental techniques; latex bilayers (circles), latex/polypropylene bilayers (diamonds) and 3D-printed PLA (triangles). Open markers correspond to finite-element (FE) simulation (see Materials and Methods for more details on the experimental techniques and the simulation).
    (b) Examples of the observed morphologies; toroidal (top), intermediate (center) and faceted (bottom) for latex bilayers (left), 3D-printed PLA (center) and FE simulations (right). See
    }
    \label{fig2}
\end{figure}    

\section{Geometry and mechanics of ribbons with geodesic and radial reference curvature}
We focus on the specific case presented in \figref{fig1}a., where a curved ribbon of geodesic curvature $\kappa_g \equiv1/R$, width $w$ and thickness $t$, with $R> w \gg t$ has a unidirectional reference curvature $\kappa$ imposed perpendicular to its midline.
The ribbon is parametrized in polar coordinates by $R-w/2 \le R+u \le R+w/2$  and $ -\theta_0/2 \le\theta \le \theta_0/2$. It appears as a slight modification of the classical bilayer structure, since the geodesic curvature $\kappa_g$ does not vanish. The reference metric and curvature tensor of the ribbon read:
\begin{equation}
    \ab = \smatrixII{1}{0}{(R+u)^2},\quad \bb=\smatrixII{\kappa}{0}{0}
    \label{eq:abbb}
\end{equation}
%Such a ribbon has thus an Euclidean reference metric and a locally unidirectional reference curvature, such that Gauss equation $\Kb=\det\bb/\det\ab=0$ is satisfied (where $\Kb$ is the reference Gaussian curvature). 
Such a reference geometry clearly obeys Gauss equation, since the metric is flat and the reference curvature is unidirectional. 
Considering now the MCP equations \ref{eq:MCP}, the first equation is trivial, but the second one reads $ 0 = -(R+u)^2 \kappa $, which is inconsistent whenever the radial reference curvature $\kappa$ is non-zero. The reference metric and curvature tensors are thus incompatible, resulting in geometrical frustration.

We use two different experimental methods to realize and quantitatively study such ribbons: latex bilayers~\cite{Armon11} (or hybrid latex and polypropylene sheets) and heat-actuated 3D-printed shape-memory polymers~\cite{an18,gu19}. In latex bilayers, one sheet is radially cut in fine stripes and radially stretched. A passive sheet (of latex or polypropylene) is glued on the stretched membrane. A curved ribbon is then cut from the membrane (see Materials and methods). 3D-printed polymers (PolyLactic Acid (PLA) filaments) uniaxially shrink along printed filaments, once the printed sheet is heated above glass temperature (see Materials and Methods and~\cite{van2017programming,an18,gu19}).
In our ribbons, the first layer is thick and filled with an isotropic filling (see Materials and Methods) and is thus mostly passive, whereas the other layer is made of thin radial lines which contract when heated, inducing the radial reference curvature.
Finite-elements simulations are also performed using an in-house code that minimizes equation \Eqref{eq:energy} given the reference geometry in equation \Eqref{eq:abbb} to find the shape of the ribbon (see Materials and Methods for more details).

In both experiments and simulations, the ribbons exhibit a rich morphology typical of frustrated ribbons depending on the four geometrical parameters (geodesic radius $R$, width $w$, thickness $t$, and radial reference curvature $\kappa$).
Three qualitatively different shapes are experimentally and numerically observed and plotted in a phase diagram (Fig.~2a): smooth toroidal shapes (in blue, Fig. 2b, top), toroidal shapes with localized defects (in orange, Fig. 2b, center) and polygonal shapes made of piecewise-tubular regions connected by corners ( in green Fig.~2b, bottom).  
The toroidal shapes exhibit an increase in their geodesic curvature, resulting in the coiling of the ribbon (Fig.~2b, top); in the following, we term this effect as \emph{overcurvature}.
When closed into a ring, this effect drives these ribbons to buckle  out-of-plane  (Fig.~3b).

\begin{figure*}[ht!]
    \centering
    \includegraphics[width=0.9\linewidth]{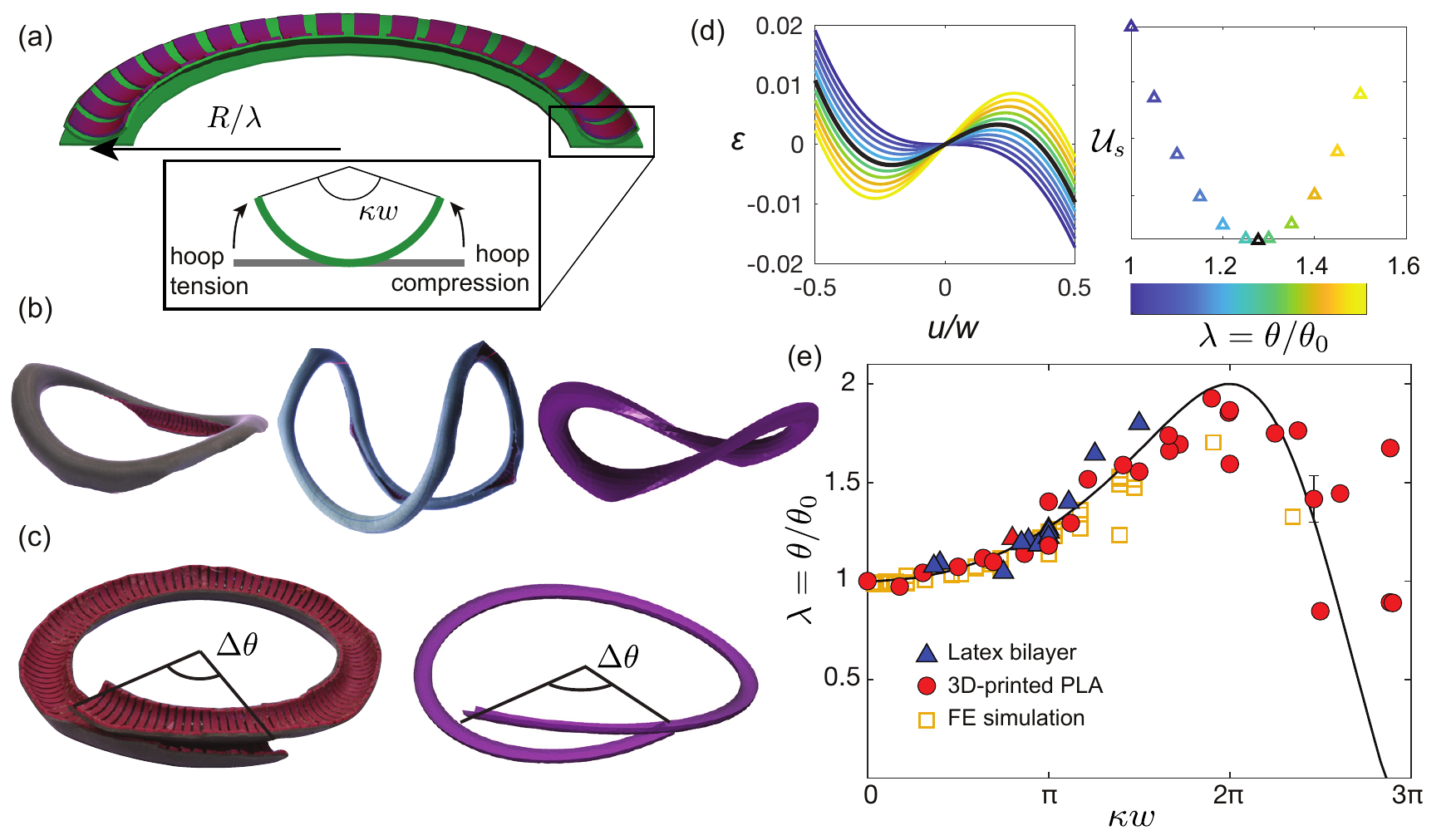}
    \caption{{\bf Overcurvature and buckling of bending-dominated ribbons}. 
    (a) In this regime, ribbons adopt a toroidal shape, inducing hoop tension and compression in the inner and outer part respectively. 
    (b) When closed, the ribbon outline buckles out of plane, with an amplitude depending on the total radial rotation angle $\kappa w$; experiments (left and center) and simulation (right) of buckled tori.  
    (c) Open ribbons remain planar and overcurve by a factor $\lambda=\theta/\theta_0$. 
    (d) (Left) Azimuthal strain profile across the width of the ribbon for various overcurvature factors $\lambda$ (with $\kappa w=\pi$, $R/w=10$). Coiling induces a reduction of the strains in the ribbon, until an optimal configuration (black line) is reached. (Right) Corresponding stretching energy as a function of $\lambda$.
    (e) Overcurvature factor $\lambda$ as a function of the total radial rotation angle $\kappa w$. Latex bilayers (circles), 3D-printed PLA (triangles) and numerical simulations (open squares) fall on the analytic curve given by equation \ref{eq:lambda}.
  }
    \label{fig:overcurve}
\end{figure*}
\begin{figure*}[ht!]
        \centering
        \includegraphics[width=0.9\linewidth]{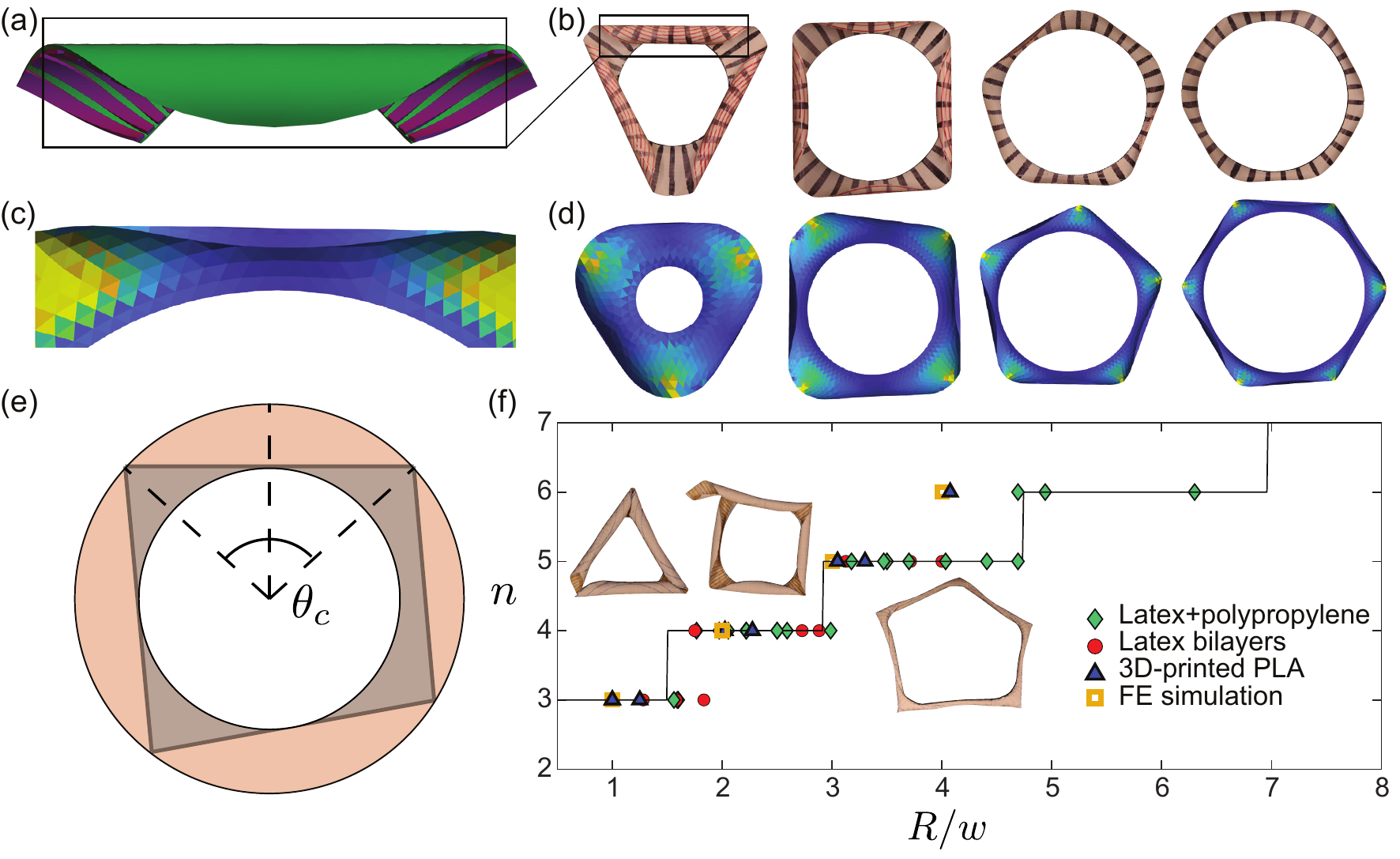}
        \caption{{\bf Faceted stretching-dominated ribbons}. 
        (a) Sketch of the locally tubular configuration observed in experiments. 
        (b) Closed ribbons (made here of polypropylene/latex bilayers) exhibit polygonal shapes, the number of sides corresponding to the number of locally tubular portions. When the width is gradually reduced, the number of sides increases. 
        (c) Finite element simulation of a portion of a curved stretching-dominated ribbon. The color codes for the bending energy density: it increases with the angle $\theta$ (as predicted by equation~\ref{eq:ub}), except along the inner boundary. 
        (d) Simulation of full structures,  with clear bending energy focusing at the junction between tubular sections. 
        (e) Geometrical rule to predict the typical angle extension $\theta_c$ of the sides and hence the number of sides.
        (f) Number of sides $n$ as a function of the slenderness ratio $R/w$ of the curved ribbons. Simulations and experimental data (pictures of some experiments in inset) are quantitatively predicted by the simple geometric construction (Equation \ref{eq:theta}).
        }
        \label{fig:polygones}
\end{figure*}

\subsection{Bending dominated regime}
Let us first investigate the blue region in Fig.~2a. The out-of-plane buckling of closed rings (Fig. \ref{fig:overcurve}b) experimentally observed, is reminiscent of curved-fold origami~\cite{dias12}, growing curved rods~\cite{moulton13} or inflated rings~\cite{siefert2019programming}. When the rings are cut, ribbons remain in-plane but exhibit a pronounced overcurvature (Fig.~3c).
Qualitatively, the radial bending of the ribbon, as shown in Fig.~3a, clearly induces negative, and positive Gaussian curvatures on the inner and outer parts of the toroidal shape adopted by the ribbon, respectively.

This configuration, although not satisfying neither the reference metric nor the curvature, is surprisingly observed experimentally in a regime of sufficiently small $\kappa w$ and sufficiently large $Rt\kappa^2$.
In the case of previously studied Gauss-frustrated sheets, the bending-dominated regime indeed corresponds to shapes obeying $\bac=\bb$, which is clearly not the case here: the radial component of the curvature tensor does obey its reference counterpart, but not the azimuthal one.

We now wish to characterize the overcurvature observed experimentally and estimate the elastic energy scaling associated with these toroidal configurations.

\subsubsection{Overcurvature characterization}
Assuming that both the center line and radial lines of the ribbon remain unstretched (to prevent significant stretching \cite{grossman16}), we postulate the following parametrization for the torus:
\begin{equation}
    \vect{x}(u,\theta)=\vecIII
    {\Brk{R/\lambda+\kappa^{-1}\sin\kappa u}\cos(\lambda\theta)}
    {\Brk{R/\lambda+\kappa^{-1}\sin\kappa u}\sin(\lambda\theta)}
    {\kappa^{-1}\cos\kappa u}
\end{equation}
where $\lambda$, which measures the overcurvature in the ribbon ($\lambda=1$ implies no over curving), is the only free parameter. 

The metric tensor of such configuration reads:
\begin{equation}
    \aac = \smatrixII{1}{0}{\brk{R+\lambda \kappa ^{-1} \sin \kappa u}^2},
\end{equation}
and its principle curvatures are given by:
\begin{equation}
     \kappa_1=\kappa, \quad
     \kappa_2=\lambda \kappa \sin \kappa u\abs{R \kappa +\lambda \sin \kappa u }^{-1}
     \label{eq:curvatures}
\end{equation}
Note that $\lambda\to0$ corresponds to the cylindrical configuration satisfying $\bac=\bb$.
The azimuthal strain in the small strain limit ($R\gg w$) reads: $\varepsilon(u)=\left(\lambda\kappa^{-1}\sin[\kappa u]- u\right)/R$. 

Minimizing the stretching energy $U_S=\frac{1}{w}\int_{-w/2}^{w/2}\varepsilon^2(u)\mathrm{d}u$ with respect to $\lambda$, we get the following expression for the overcurvature:
\begin{equation}
 \lambda= \dfrac{4\sin(\kappa w/2)-2\kappa w\cos(\kappa w/2)}{\kappa w-\sin(\kappa w)}
 \label{eq:lambda}
\end{equation}
More intuitively, considering the radial bending of the structure without overcurvature, azimuthal stretching and compression appear respectively at the inner and outer parts of the ribbon (Fig.~\ref{fig:overcurve}a).
Such a strain (and thus stress) distribution creates a torque in the structure, resulting in overcurvature.
The hoop strain profile along the width (shown in Fig.~\ref{fig:overcurve}d) evolves with the overcurvature $\lambda$. The ribbon naturally selects one given overcurvature $\lambda_{opt}$, for which the profile is optimal and the stretching energy is minimized (black line in Fig.~\ref{fig:overcurve}d).
The overcurvature monotonically increases with $\kappa w$ and reaches its maximal value 2 when the structure self closes and forms a complete torus. 
It then abruptly drops for larger values of the total radial rotation angle $\kappa w$ and the structure even undercurves when $\kappa w$ approaches $3\pi$. 
This theoretical curve remains valid as long as the geodesic radius $R$ is substantially larger than the width $w$, such that compressive and tensile strains (in the outer resp. inner part of the ribbon) may be considered symmetric. 
This analytical prediction is in good agreement with the overcurvature, measured in the experiments and simulations (Fig.~\ref{fig:overcurve}d) in the bending-dominated regime, as long as the total radial rotation angle $\kappa w$ is smaller than $2\pi$.
Indeed, above this critical value, overlap and self contact appear and friction plays an important role, making the experiments more sensitive.

Based on the strain and curvature expressions found in the last paragraph, one may now estimate the scaling of the energy, depending on the value of the total rotation angle $\kappa w$ (see the inset in Fig.\ref{fig:overcurve}a for the geometric interpretation of this angle).
 
\subsubsection{Energy scaling for $\kappa w <1$}
When the angle $\kappa w$ is small, the typical azimuthal curvature (equation \ref{eq:curvatures}) simplifies to $\inv{R}\kappa w$ (we assume here $\lambda=1$ for the sake of simplicity, and one may show that different values of the overcurvature do not affect the energy scaling).
As shown before, the inner and outer boundaries of the ribbon are under hoop tension and compression respectively,  with a typical strain $\epsilon_S \approx \kappa^2 w^3 \inv{R}$.
The energy density of the ribbon scales as
\begin{equation}
    \mathcal{U} \sim w^6  \kappa^4 \invs{R}+ t^2 R^{-2} (\kappa w)^2 
    \label{eq:toroid}
\end{equation}
One might expect, as in Gauss-frustrated sheets, that the  structure would  obey the reference curvature to cancel the bending content in the bending-dominated regime. Following the uniaxial reference curvature $\bb$ however requires the straightening of the ribbon in order to bend into a cylindrical shape: the subsequent stretching energy scales as $\mathcal{U}_S \sim w^2 \invs{R}$. This energy appears to be asymptotically larger than both energetic terms in equation~\ref{eq:toroid} explaining why such a configuration is not favored and thus not experimentally observed. Comparing both terms in equation~\ref{eq:toroid}, the stretching term is larger than the bending one as long as $\kappa w>t/w$. This is always the case in interesting structures, since $t\ll w$; an angle $\kappa w \ll 1$ rad would result in a mostly flat and barely frustrated ribbon.  

\subsubsection{Energy scaling for $\kappa w \sim 1$}
When the angle $\kappa w$ is large , the typical strain in a toroidal configuration simply reads $\epsilon_S \approx w\inv{R}$ as in the case of a straight tubular configuration. The typical azimuthal curvature coincides with the geodesic curvature $\lambda/R$. The energy thus reads 
\begin{equation}
  \mathcal{U} \sim w^2 \invs{R}+t^2 R^{-2} 
  \label{eq:toroid2}
\end{equation}
In this case again, the bending term is negligible compared to the stretching one (since $w\gg t$), and the energy scalings of the toroidal and tubular configurations are equivalent; the value of the overcurvature $\lambda$ does not change the scaling of the energy, only the prefactor.

\subsection{Stretching dominated regime}
In the wide stretching-dominated limit (green region in Fig.~2a), the selected shape has to be found among embeddings of the reference metric, which is in our case Euclidean. That is, we seek to find a developable surface (i.e. a surface with zero Gaussian curvature) that minimizes the bending energy. In this regime, we experimentally observe the formation of straight tubular shapes connected by unbent corners (Fig.~4ab). 
Considering a ribbon rolled into a cylinder of radius $\kappa^{-1}$ along the mean radial direction of the ribbon (Fig.~4a), the curvature mismatch and hence the bending energy density increase with the angle $\theta$ (Fig.~4cd).
Switching to Cartesian coordinates $(x,y)$, where the direction ${\bf e}_y$ is chosen to be the direction of curvature, the actual and reference curvature tensors indeed respectively read:
\begin{equation}
    \bb = \smatrixII{\kappa\sin^2\theta}{\kappa\cos\theta\sin\theta}{\kappa\cos^2\theta}
    ,\quad \quad 
    \bac= \smatrixII{0}{0}{\kappa}
\end{equation}
where $\theta$ is the angle made between ${\bf e}_y$ and ${\bf e}_r$.
The local dimensionless bending energy density reads:
\begin{equation}
    u_B=t^2\left[(1-\nu)\kappa^2\sin^2\theta+4\nu\kappa^2\sin^4\theta\right]
    \label{eq:ub}
\end{equation}
This expression is a growing function of $\theta$. For small angles, this energy is lower  than the energy density required to flatten the structure, for which $u_B=t^2\kappa^2$.
The rolling of the ribbon onto a tube is thus energetically favored. The smaller the extent of the tubular portion (in terms of geodesic angle $\theta$), the smaller the bending energy density. 
Hence, a thin ribbon tends to adopt a piecewise-tubular shape to approximate the reference radial curvature, exhibiting polygonal shapes reminiscent of geometrically confined bilayer shells~\cite{stein19} (Fig.~4bd).
The minimum possible geodesic curvature angle $\theta_c$ for unidirectional bending that can be connected without stretching can be obtained through a simple geometric construction (Fig.~4e): the region delimited by the intersection of the outer boundary and the tangent to the inner boundary, can indeed freely bend independently from the other regions. Its angle extension reads:
\begin{equation}
    \theta_c=2\arccos{\frac{2R-w}{2R+w}}
    \label{eq:theta}
\end{equation}
This simple rule predicts quantitatively the number of facets in stretching-dominated samples, both for opened and closed ones (Fig.~4f).
Remarkably, this relation depends only on the lateral geometry of the ribbon (i.e. the aspect ratio $R/w$) and not on the thickness $t$ nor the reference curvature $\kappa$.

Note that a region of finite arclength (highlighted in shades in Fig.~4e) must radially unbend at the junction between these tubular portions.
The bending energy density thus typically scales as 
\begin{equation}
 \mathcal{U}_B \sim t^2 \kappa^2  
 \label{eq:bending}
\end{equation}
In these polygonal configurations, the stretching energy density vanishes almost everywhere, and most of the bending energy is localized at the corners separating two tubular sections (\figref{fig:polygones}d). Some stretching occurs at these corners to regularize the otherwise diverging azimuthal bending energy. Although being free of external constraints and having a smooth reference geometry, these ribbons remarkably exhibit stress focusing~\cite{witten2007stress}.

\subsection{Shape transitions}
The transition between both regimes may be estimated by balancing energies in both asymptotic regimes. As the bending term in equations \ref{eq:toroid} and \ref{eq:toroid2} is much smaller than the stretching one, the typical transition between the bending and stretching dominated regimes may be obtained by equating the bending energy in the stretching-dominated configuration (equation \ref{eq:bending}) and the stretching energy of the toroidal configuration (equations \ref{eq:toroid} and \ref{eq:toroid2}). We get the following characteristic transition widths: 
\begin{equation}
    \tilde{w_1}\sim \left(tR/\kappa\right)^{1/3}
    \label{eq:trans1}
\end{equation}
for $\kappa w <1$,
and 
\begin{equation}
    \tilde{w}\sim tR\kappa
    \label{eq:trans2}
\end{equation}
for $\kappa w \sim1$.

Another intuitive way of thinking of this transition is to consider the compressive azimuthal strain $\epsilon_S$ at the outer boundary in the torus configuration and estimate when this strain reaches the buckling threshold of a cylindrical shell of curvature $\kappa$ and thickness $t$, which typically reads $\epsilon_c\sim t\kappa$.

In order to test these scaling laws, we plot in Fig.~\ref{fig2}a the total radial rotation $\kappa w$ as a function of the dimensionless quantity $Rt\kappa^2$. 
%This quantity is the product of $\kappa t$ which estimates the typical radial prestrain in the structure and $\kappa R$ which is the source of geometrical incompatibility. 
As expected by equation \ref{eq:trans1}, the transition lies along a line of slope 3 for a relatively small total radial rotational angle $\kappa w$. For large $\kappa w$, the slope of the transition becomes 1, as predicted by equation \ref{eq:trans2}. 
These scalings hold for different experimental realizations and simulations. Note that for such MCP-frustrated ribbons, the scaling of the transition between bending and stretching dominated regimes is different from Gaussian-based frustration (typically $\tilde{w} \sim \sqrt{t/\kappa}$).
%The transition between bending and stretching-dominated regimes obeys in this bending-based frustrated structures different scaling laws from Gaussian-based frustrated ribbons. They are robust against different experimental realizations and simulations.

The orange region in Fig. \ref{fig2}a appears to be a mixed intermediate regime: overcurvature is observed, indicating stretching, and defects are present, but do not obey the geometric rule observed in the stretching-dominated regime. At the inner part of the ribbon, a boundary layer with Gaussian curvature is observed (see Fig.~4c) and extends over a typical width $\tilde{w}$, the transition width derived in Eq. \ref{eq:trans1}. 
This scaling is once again different from that of boundary layers in Gauss-frustrated ribbons~\cite{efrati2009buckling} where $\tilde{w} \propto \sqrt{t}$.

\section{Discussion and Conclusion}
In summary, we presented a novel class of geometric frustration based on the spatial variation of the reference curvature tensor, in which Gauss equation is satisfied but not the Mainardi-Codazzi-Peterson equations. 
We focused on one specific model case, where ribbons with geodesic curvature have a constant reference curvature along their width.
This system, although implying a minimal modification from the well-studied unidirectional bilayer structures, exhibits a rich morphology and new scaling laws for the shape transition.

Such ribbons share a striking resemblance with curve-fold origami~\cite{dias12}. Its prescribed azimuthal fold induces frustration in the structure, leading to a competition between the bending of the sheet and the opening of the fold. Overcurvature and out-of-plane buckling are similarly observed in open and closed structures, respectively. We note, however, that the localization of the radial curvature on the fold plays a significant role: it enables the structure to smoothly obey the total radial rotation without inducing Gaussian curvature. Therefore, the configurations observed are stretch-free, and no shape transition manifests. Moreover, distributing the curvature by introducing several consecutive azimuthal mountain folds is prohibited as it would result in stretching.

In a wider perspective, this work raises several fundamental questions and new theoretical developments are needed to incorporate such MCP-frustration in the general theory of frustrated shells.
In particular, Gauss-frustration may lead to anomalous mechanics~\cite{guest11,levin16}. Hence, a natural extension of this work is to study the energy landscape and the mechanical properties of MCP-frustrated ribbons.
Being as general as Gauss-frustration, we anticipate MCP-frustration to appear in many biological and responsive systems and to play a significant role in future devices and applications. The spontaneous symmetry breaking and stress-focusing could be harnessed in order to obtain specific shapes or functions from flat ribbons.

\section*{Materials and Methods}
\begin{figure}
    \centering
    \includegraphics[width=1\linewidth]{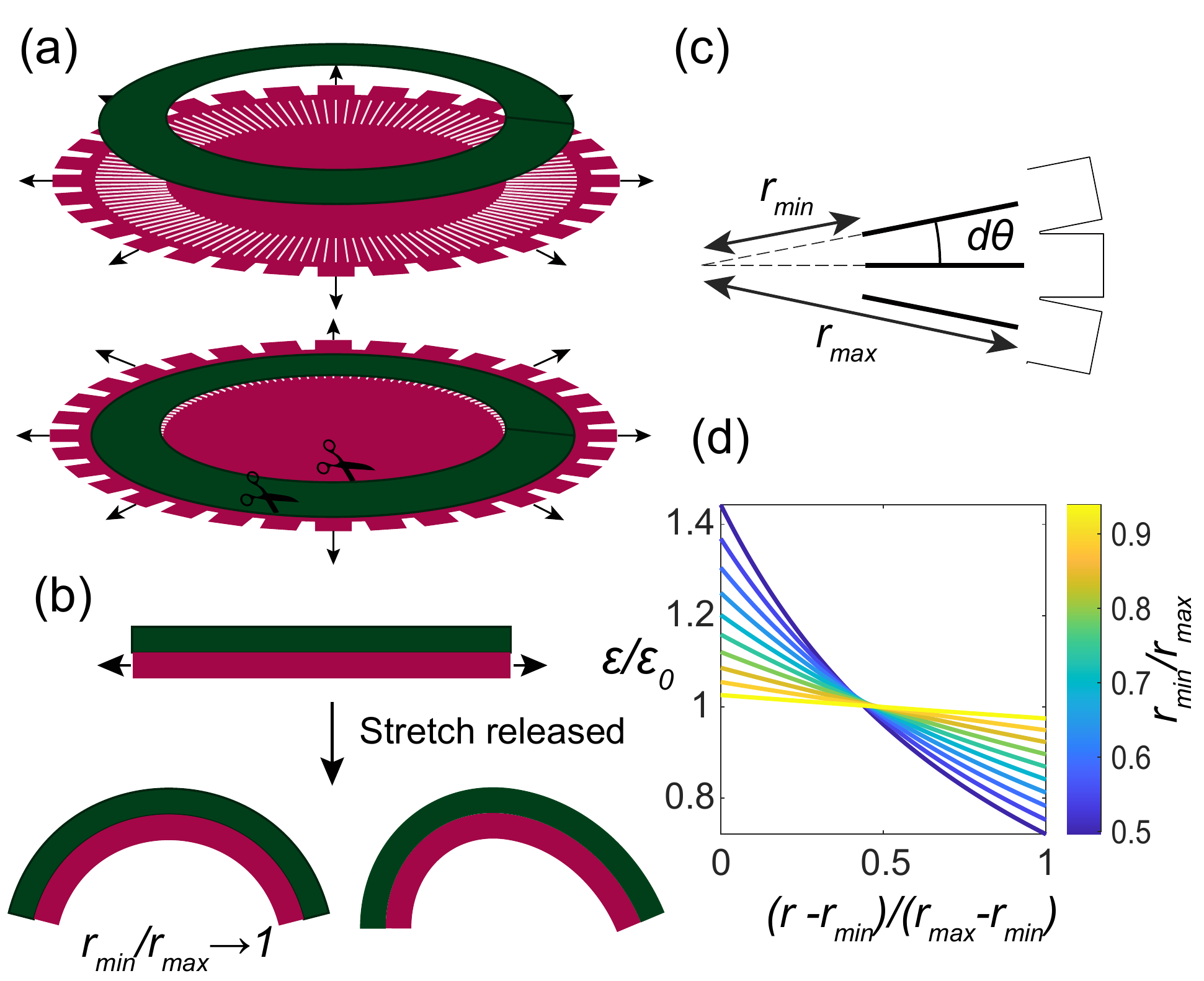}
    \caption{{\bf Stretched latex bilayers}. 
    (a) A latex sheet is stretched radially by a pre-determined factor. The sheet is then glued to an unstretched sheet made of latex or polypropylene. The radial cuts release the hoop stress resulting from the Poisson effect, inducing a quasi-radial stretch. 
    (b) When cut into a curved ribbon and thus released, a radial reference curvature is imposed to the structure. This curvature is however not exactly constant. Depending on the the size of the radial cuts, and more specifically on the ratio $r_{min}/r_{max}$ (c), the radial strain, and thus the target curvature, vary as a function of the radial coordinate $r$ (d). This effect is however neglected in our study.
    }
    \label{fig:latex}
\end{figure}

\paragraph*{Stretched  Latex bilayers}
A latex sheet is radially cut with a laser cutter (cuts extending between $r_{min}$ and $r_{max}$) every small angle $d\theta$ in order to form radial strips (\figref{fig:latex}a, top). The latex sheet is then radially stretched and fixed on a circular frame. A thin layer of latex-based glue (Free chack II from Butterfly) is then spread uniformly on the sheet and a second unstretched layer is glued on the stretched membrane. After curing, a ring of mid-radius $R$ and width $w$ is cut from the sheet, ensuring that it contains only the radially cut portion of the stretched sheet. Once released, the stretched strips induce a typical radial bilayer curvature $\kappa$ with magnitude proportional to the imposed radial strain $\varepsilon_0$ (\figref{fig:latex}b). Rigorously, the imposed radial strain is not constant along the radial coordinate, but these variations are small and we neglect them in this work (see \figref{fig:latex}d, supplementary material). 
The strips are indeed not rectangular and have a width $r_{min} \dif\theta$ at their base and $r_{max}d\theta$ at their outer boundary (\figref{fig:latex}c). The constant force $F$ imposed on the stretched strips induces a gradient in stresses $\sigma=F/(t r \, \dif\theta)$ and thus a varying strain $\varepsilon=\frac{F}{E t r \, \dif\theta}$ in the framework of linear elasticity. Experimentally, the displacement $\delta$ is imposed. It corresponds to the integrated strain over the full length of the strip:  $\delta=\varepsilon_0(r_{max}-r_{min})=\int_{r_{min}}^{r_{max}}\varepsilon \mathrm{d}r$. Combining last two equations, the strain along the strip reads:
\begin{equation}
    \varepsilon(r)=\varepsilon_0\frac{r_{max}-r_{min}}{r\ln{r_{max}/r_{min}}}
\end{equation}
where $\varepsilon_0=\delta/(r_{max}-r_{min})$ is the average imposed strain. The strain evolution in the stretched strips are plotted in \figref{fig:latex}d) for various values $r_{min}/r_{max}$. When $r_{min}/r_{max}\to 1$, i.e., when the cuts are small compared to the radius $r_{min}$, the strain is almost constant. From Timoshenko~\cite{timoshenko25}, we know that the reference curvature in bilayers scales linearly with the strain difference $\varepsilon$. $r_{min}/r_{max}\to 1$ means thus constant radial reference curvature (\figref{fig:latex}b, left), whereas the curvature significantly varies for smaller values of $r_{min}/r_{max}$ (\figref{fig:latex}b, right).

In this article, we neglect the radial variation of the reference curvature in stretched latex bilayers described above, since we mostly consider the case where $R\gg w$. In order to estimate the mean curvature when it is not constant, we measure the total radial rotation angle and divide it by the width $w$.

\begin{figure}
    \centering
    \includegraphics[width=1\linewidth]{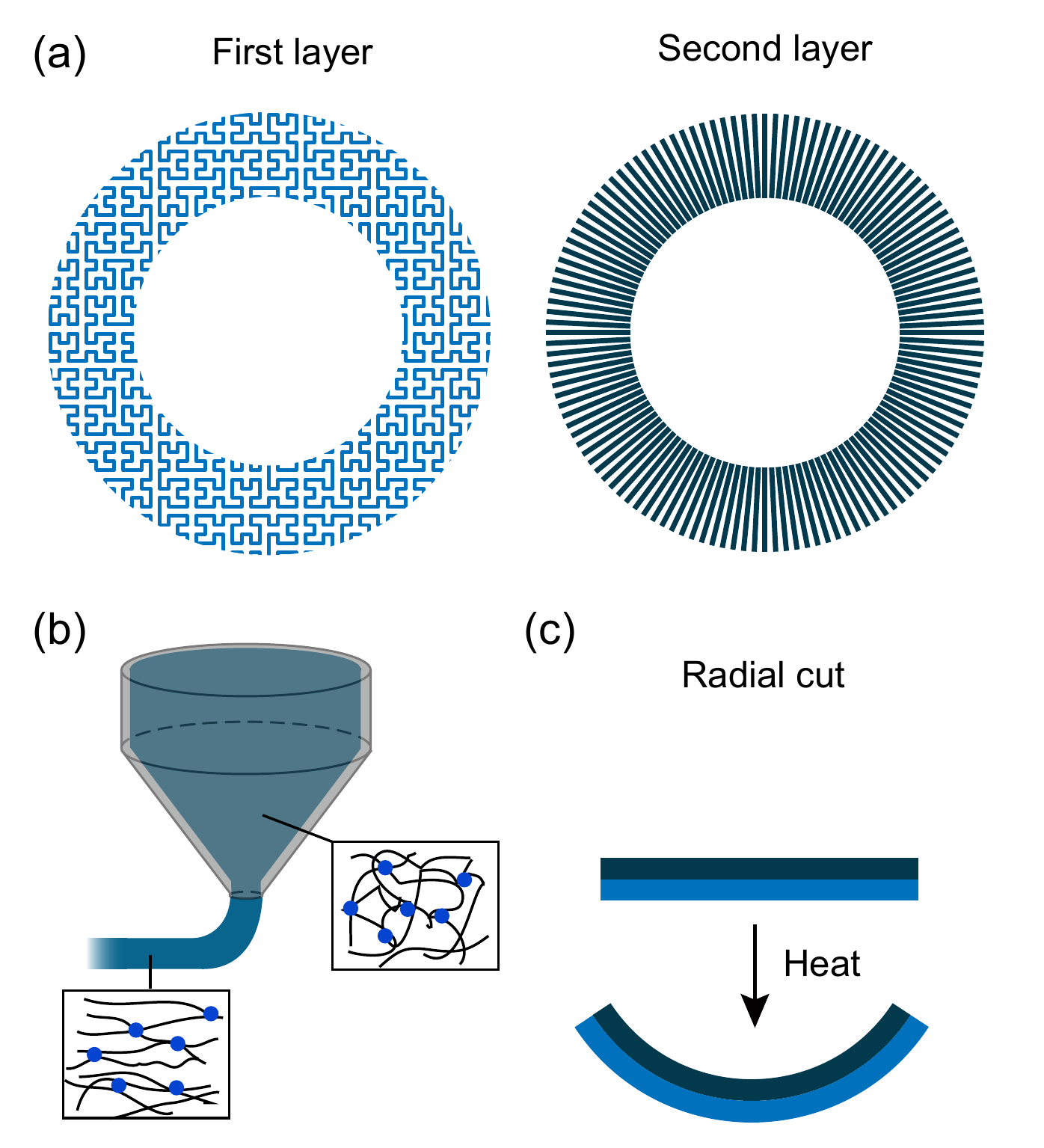}
    \caption{{\bf 3D-printed thermoplastic ribbons}.
    (a) Radial lines (second layer) are printed on top of a layer printed along a Hilbert curve (first layer). 
    (b) Long polymeric chains straighten in the nozzle due to the extensional flow. Cooling down, this elongated configuration results in residual stresses along the printing direction. When heated above the glass transition temperature, the printed thermoplastic material shrinks in the printing direction to release the residual stresses.
    (c) While the top layer shrinks radially, the bottom layer is mostly passive due to its thickness and the isotropic pattern, inducing thus a radial reference curvature. 
    }
    \label{fig:3Dprinting}
\end{figure}

\paragraph*{Shape memory 3D-printed PLA}
Ribbons are printed with a Fused Filament Fabrication 3D-printer (Prusa MK3S, nozzle diameter $0.4~mm$, PLA filament diameter $1.75~mm$, bed temperature $60^{\circ}C$, nozzle temperature $200^{\circ}C$ and printing speed $50~mm/s$).  
During the printing process, the PLA chains straighten in the nozzle due to the extensional flow. When the layer thickness is small the material cools down quickly and the chains are frozen in the entropically unfavourable elongated configuration, leading to residual stresses. When the material is heated up above the glass transition temperature, the material relaxes the stresses, leading to a contraction along the printing direction.
As shown in Refs.~\cite{van2017programming,an18,gu19}, the printed layer thickness $t_0$ of each layer is the critical parameter to induce large residual stresses along the printing direction. Seeking a large bilayer effect, the first layer is thus printed with a large thickness $t_0=0.15~mm$, with a Hilbert curve filling, resulting in an isotropic -0.05 strain (\figref{fig:3Dprinting}). The next layers are made of thin ( typically $t_0=0.05mm$) radial lines.
After printing, the samples are immersed in a bath of hot water (at $80^\circ C$) and deform to relax the flow-induced internal stresses. While the first layer is essentially passive, the next layers contract radially (with typical radial strains ranging from -0.1 to -0.3) inducing a radial reference curvature $\kappa$ proportional to the contraction strain.
\paragraph*{Numerical methods}
In order to test our results and predictions numerically we used a finite element code developed in-house.

We tested samples of $\theta_0 = \pi$ (a single turn), $w = 25,50$, $w \le R \le 400$, $0.05\le t\le1$ and $0.1 \le \kappa w \le 4 \pi$.
We also simulated ribbons with a topology of a closed ring, especially in the stretching dominated regime.
We input our code a triangulated domain in coordinate space (i.e. $u$ and $\theta$) corresponding to the sample tested along with the reference tensors, $\ab$ and $\bb$, and an initial configuration $\vec{x}(u,\theta)$.
In order to capture the focusing of the curvature, we used a find mesh of over 3500 triangles.
The triangulation was made such that triangle areas will be uniform in the initial configuration.
The code minimize the two-dimensional energy functional in \Eqref{eq:energy} by estimating the actual metric and curvature of the triangulated configuration and evolving it via gradient-descent.
\chng{Although both numerical and experimental results represent only a local minimum of the energy, their similarity and their agreement with the theory demonstrate the robustness of our procedure.}
Our numerical code does not penalize \chng{intersection, which generally might affect the configuration of frustrated sheets}.
Nevertheless, we expect such interaction to play limited role in the shape selection.

The initial configuration was chosen to be a flat arc with width $w$ and geodesic curvature $\inv{R}$.
To test for numerical convergence we tested other initial configuration: toroidal configurations, and a flat arcs with over curvature similar to $\lambda(\kappa w)$.
In the majority of parameter-space we found no significant changes between these initial conditions.
The most delicate part was the numerical estimation of the number of facets, $n$.
This case involves a nucleation problem, hence once $n_0$ nodes are spontaneously generated this number only increases during the minimization process.
Therefore to test for this number we used initial conditions of a flat arc with a periodic perturbation of $n_0 \in \BRK{3,...,7}$.
Then we looked for the final configuration with the minimal energy and regarded it as the energy minimizer.

\section*{Acknowledgments}
This research was supported by the USA-Israel binational science foundation, Grant No. 2014310.
I. L. is grateful to the Azrieli Foundation for the award of an Azrieli Fellowship. E. S. acknowledges support from the Lady Davis Fellowship Trust.

\bibliography{references.bib}

\end{document}